\documentstyle[prl,aps,preprint,tighten,floats,epsf,rotate]{revtex}

\newbox\rotbox

\begin{document}

\preprint{\vbox{\it Submitted to Phys.\ Lett. B
                                          \null\hfill\rm OSU-PP-95-10 \\
                                          \null\hfill\rm TRI-PP-95-66 \\
                                      \null\hfill\rm UW-DOE/ER/40427-18-N95 \\
                                         \null\hfill\rm nucl-th/9511007}}
%
\title{\huge New QCD Sum Rules\\ for\\ Nucleons in Nuclear Matter}
\author{\sc R. J. Furnstahl,$^{\rm 1}$ Xuemin Jin,$^{\rm 2}$\thanks{%
Address after September 1, 1996: Center for Theoretical Physics,
Laboratory for Nuclear Science and Department of Physics,
Massachusetts Institute of Technology, Cambridge, 
Massachusetts 02139, USA} 
and Derek B. Leinweber$^{\rm 2,3}$} 
\address{$^{\rm 1}$Department of Physics, The Ohio State
              University, Columbus, Ohio 43210, USA} 
\address{$^{\rm 2}$TRIUMF, 4004 Wesbrook Mall, Vancouver, B.C,
                V6T 2A3, Canada} 
\address{$^{\rm 3}$Department of Physics, Box 351560, University of Washington,
                Seattle WA 98195, USA} 
\date{\today}
\maketitle

\begin{abstract}
Two new QCD sum rules for nucleons in nuclear matter are obtained from
a mixed correlator of spin-1/2 and spin-3/2 interpolating
fields. These new sum rules, which are insensitive to the poorly known
four-quark condensates, provide additional information on the nucleon
scalar self-energy.  These new sum rules are analyzed along with
previous spin-1/2 interpolator-based sum rules which are also
insensitive to the poorly known four-quark condensates.  The analysis
indicates consistency with the expectations of relativistic nuclear
phenomenology at nuclear matter saturation density.  However, a weaker
density dependence near saturation is suggested.  Using previous
estimates of in-medium condensate uncertainties, we find $M^* =
0.64^{+0.13}_{-0.09}$ GeV and $\Sigma_v = 0.29^{+0.06}_{-0.10}$ GeV at
nuclear matter saturation density.  \\
\end{abstract}
%

\newpage
\narrowtext

Understanding the observed properties of hadrons and nuclei from
quantum chromodynamics (QCD) is a principal goal of nuclear theorists.
The QCD sum-rule approach \cite{shifman1} is a particularly useful
method for connecting the properties of QCD to observed nuclear
phenomena\cite{cohen1,furnstahl1,jin1,jin2,cohenr,%
drukarev91,hatsuda91,henley93,schafer93}.  Recent progress in
understanding the origin of the large and canceling isoscalar
Lorentz-scalar and -vector self-energies for propagating nucleons in
nuclear matter has been made via the analysis of QCD sum rules
generalized to finite nucleon density
\cite{cohen1,furnstahl1,jin1,jin2,cohenr}.  These large self-energies
are central to the success of relativistic nuclear
phenomenology\cite{serot1}.

However, the previous sum-rule predictions for the scalar self-energy
are sensitive to the density dependence of chirally-even dimension-six
four-quark condensates.  In-medium factorization expresses these
chirally-even operators in terms of the square of a chirally-odd
operator whose density dependence is likely to be very different.  As
such, the density dependence of these problematic four-quark operators
is generally unknown \cite{furnstahl1,jin2,cohenr}.  

   There are various ways to clarify the situation.  One approach
attempts to better determine the density dependence of the four-quark
condensates via modeling \cite{celenza1}.  One may also use other
independent information to constrain the in-medium four-quark
condensates \cite{johnson95}.  An alternative approach, which is
adopted here, is to derive new QCD sum rules that are insensitive to
the four-quark condensates.

In this work, we obtain two new sum rules from a mixed correlator of 
generalized spin--1/2 and spin--$3/2$ interpolating fields.  
The spin--$1/2$ states remain projected, and one generates 
additional sum rules for the nucleon scalar self-energy that are 
insensitive to the four-quark condensates.  Our hope is that these 
new sum rules, along with previous spin--1/2 interpolator-based sum
rules which are insensitive to the problematic four-quark 
condensates \cite{furnstahl1,jin2,cohenr}, will allow a better
determination of the nucleon self-energies in nuclear matter.

   The finite-density QCD sum-rule approach focuses on a correlation
function of interpolating fields carrying the quantum numbers of the
hadron of interest. The correlation function is evaluated in the
ground state of nuclear matter instead of the QCD vacuum.  The
appearance of an additional four-vector at finite density, the
four-velocity of the nuclear medium, leads to additional invariant
functions relative to the vacuum case \cite{cohen1,furnstahl1,%
jin1,jin2,cohenr,jinm}.  In the rest frame of the medium, the analytic
properties of the various invariant functions can be studied through
Lehman representations in energy.  The quasi-nucleon excitations
(i.e., the quasiparticle excitations with nucleon quantum numbers) are
characterized by the discontinuities of the invariant functions across
the real axis, which are used to identify the on-shell self-energies.
A representation of the correlation function can be obtained by
introducing a simple phenomenological Ansatz for these spectral
densities.

  On the other hand, the correlation function can be evaluated at
large space-like momenta using an operator product expansion (OPE).
This expansion requires knowledge of QCD Lagrangian parameters and
finite-density quark and gluon matrix elements (condensates).
Finite-density QCD sum rules, which relate the nucleon self-energies
in the nuclear medium to these QCD inputs, are obtained by equating
the two different representations using appropriately weighted
integrals \cite{furnstahl1,jin2,cohenr}.

Consider the correlation function defined by
\begin{equation}
\Pi_{\mu\nu}^{12}(q)\equiv i\int d^4 x\, e^{iq\cdot x}
\langle\Psi_0|T \left[ \chi^1_\mu(x) \, \overline{\chi}^2_\nu(0)
\right ]| \Psi_0 \rangle \, ,
\label{corr_def}
\end{equation} 
where the ground state of nuclear matter $|\Psi_0\rangle$ is
characterized by the rest-frame nucleon density $\rho_N$ and by the
four-velocity $u^\mu$; it is assumed to be invariant under parity and
time reversal except for the transformation of $u^\mu$. The
interpolating fields are taken to be \cite{leinweber1,leinweber2}
\begin{eqnarray}
\chi^1_\mu(x)&=& \gamma_\mu \gamma_5 \epsilon^{abc}\left\{
\left[u_a^T(x) \, C \, \gamma_5 \, d_b(x) \right] u_c 
+ 
\beta \left[u_a^T(x)\, C\, d_b(x)\right]\gamma_5\, u_c(x)
\right\}\, ,
\label{field-1}
\\*[7.2pt]
\chi^2_\nu(x)&=&\epsilon^{abc}\left\{
\left[u_a^T(x)\, C\, \sigma_{\rho\lambda}\, d_b(x) \right]
\sigma^{\rho\lambda} \, \gamma_\nu \, u_c(x)
-\left[u_a^T(x) \, C\, \sigma_{\rho\lambda}\, u_b(x) \right]
\sigma^{\rho\lambda} \, \gamma_\nu \, d_c(x) \right\} \, ,
\label{field-2}
\end{eqnarray}
where $T$ denotes a transpose in Dirac space, $C$ is the charge
conjugation matrix, and $\beta$ is a parameter allowing for arbitrary
mixing of the two independent spin-1/2 interpolating fields
\cite{leinweber3}.  

The correlator of $\chi^1_\mu$ and $\overline{\chi}^1_\nu$ gives the
sum rules discussed extensively in
Refs.~\cite{cohen1,furnstahl1,jin1,jin2,cohenr}. In particular, the
sum rule at the structure, $\gamma_\mu \rlap{/}q \gamma_\nu$, strongly
depends on the even-chirality four-quark condensates,
$\langle\overline{q}\, \Gamma_i\, q\, \overline{q}\, \Gamma_i\,
q\rangle_{\rho_N}$ and $\langle\overline{q}\, \Gamma_i\, \lambda_{a}\,
q\, \overline{q}\, \Gamma_i\, \lambda_{a}\, q\rangle_{\rho_N}\, ,$
where $\lambda^{a}$ are the Gell-Mann matrices and $\Gamma_i$ are any
of the 16 Dirac matrices.  The other two sum rules are insensitive to
these even-chirality four-quark condensates and are given by
\begin{eqnarray}
\lambda_1^2 \, M_N^\ast \, e^{-(E_q^2-{\bf q}^2)/M^2}
&=&- \, {7 - 2 \beta - 5 \beta^2\over 64 \pi^2} \,
   M^4 \, E_1 \, \langle\overline{q}q\rangle_{\rho_N}
\nonumber\\*
& &\null
   +{3 \left ( 1 - \beta^2 \right ) \over 64 \pi^2}\, M^2 \, E_0 \,
   \langle g_c \overline{q}\, \sigma\!\cdot\!{\cal G}q\, 
   \rangle_{\rho_N} \, L^{-14/27}
\nonumber\\*
& &\null
   -{7 - 2 \beta - 5 \beta^2\over 12} \, 
   \overline{E}_q \, \langle\overline{q}q\rangle_{\rho_N} \,
   \langle q^\dagger q\rangle_{\rho_N} 
\nonumber\\*
& &\null
   -{9 + 10 \beta - 29 \beta^2 \over 2^7 \, 3^2} \, 
   \langle\overline{q}q\rangle_{\rho_N} \,
   \langle {\alpha_s \over \pi} G^2 \rangle_{\rho_N}\, ,
\label{sum-1}
\end{eqnarray}
and
\begin{eqnarray}
\lambda_1^2 \, \Sigma_v \, e^{-(E_q^2-{\bf q}^2)/M^2}
&=&{5 + 2 \beta + 5 \beta^2 \over 48 \pi^2} \, M^4\,  E_1 \, 
\langle q^\dagger q \rangle_{\rho_N} \, L^{-4/9}
\nonumber\\*
& &\null
+{5 \left (5 + 2 \beta + 5 \beta^2 \right ) \over 72 \pi^2} \,
 \overline{E}_q \, M^2 \, E_0 \,
 \langle q^\dagger iD_0 q \rangle_{\rho_N} \, L^{-4/9}
\nonumber\\*
& &\null
-{7 + 10 \beta + 7 \beta^2 \over 192 \pi^2} \, M^2\,  E_0\,
\langle g_c q^\dagger\sigma\!\cdot\!{\cal G} q\rangle_{\rho_N} \, 
L^{-4/9}
\nonumber\\*
& &\null
+{5 + 2 \beta + 5 \beta^2 \over 8 \pi^2} \, {\bf q}^2
\left ( \langle q^\dagger iD_0 iD_0 q\rangle_{\rho_N}+{1 \over 12}
   \langle g_c q^\dagger\sigma\!\cdot\!{\cal G}q\rangle_{\rho_N}
\right ) \, L^{-4/9}
\nonumber\\*
& &\null
+{5 + 2 \beta + 5 \beta^2 \over 12} \, \overline{E}_q \, \kappa \,
\langle q^\dagger q\rangle_{\rho_N}^2 \, L^{-4/9}\ .
\label{sum-2}
\end{eqnarray}
Here $\lambda_1$ denotes the coupling of $\chi^1_\mu$ to the
quasi-nucleon state.  We have also defined
\begin{equation}
M_N^* \equiv M_N+\Sigma_s\ ,\hspace*{1cm}
E_q\equiv \Sigma_v+\sqrt{{\bf q}^2+M_N^{\ast 2}}\ ,\hspace*{1cm}
\overline{E}_q \equiv \Sigma_v-\sqrt{{\bf q}^2+M_N^{\ast 2}}\ ,
\end{equation}
where $\Sigma_s$ and $\Sigma_v$ are the scalar and vector
self-energies of the nucleon in nuclear matter, respectively.  The
anomalous dimensions of various operators have been taken into account
through the factor $L \equiv \ln(M^2/ \Lambda_{\rm
QCD}^2)/\ln(\mu^2/\Lambda_{\rm QCD}^2)$ \cite{shifman1}.  We have also
defined $E_0\equiv 1-e^{-s_0/M^2}$ and $E_1\equiv
1-e^{-s_0/M^2}\left(s_0/ M^2+1\right)$, which account for
excited-state contributions \cite{furnstahl1,jin2,cohenr,leinweber4}.

In sum rules (\ref{sum-1}) and (\ref{sum-2}), contributions
proportional to the up and down current quark masses have been
neglected as they give numerically small contributions. The
contributions of the even-chirality four-quark condensates to the sum
rule (\ref{sum-1}) are proportional to current quark masses and can
thus be neglected safely. Their contributions to (\ref{sum-2}) appear
only in the form of
$\langle\overline{q}\, \gamma_0\, q\, \overline{q}\, 
\gamma_0\, q\rangle_{\rho_N} - {1\over 4}
\langle\overline{q}\, \gamma^\mu\, q\, \overline{q}\, 
\gamma_\mu\, q\rangle_{\rho_N}$ or 
$\langle\overline{q}\, \gamma_0\, \lambda_{a}\, q\, \overline{q}\, 
\gamma_0\, \lambda_{a}\, q\rangle_{\rho_N} -{1\over 4}
\langle\overline{q}\, \gamma^\mu\, \lambda_{a}\, q\, \overline{q}\, 
\gamma_\mu\, \lambda_{a}\, q\rangle_{\rho_N}$. 
It is easy to see that these two combinations go to zero in the zero
density limit and they are proportional to $\rho_N$ in the linear
density approximation. Here we will assume in-medium factorization for
these {\it combinations}.  To explore the sensitivity of this
assumption, we have introduced the parameter $\kappa$ in (\ref{sum-2})
such that deviations from in-medium factorization $(\kappa = 1)$ may
be explored during the Monte-Carlo based uncertainty analysis.  The
remaining nonvanishing four-quark condensates, $\langle \overline{q}
\, q\, q^\dagger\, q \rangle_{\rho_N}$ and $\langle \overline{q}\,
\lambda_a\, q\, q^\dagger\, \lambda_a\, q \rangle_{\rho_N}$, have odd
chirality and vanish at zero density.  For simplicity, these
condensates are approximated by their factorized values.

   The interpolating field $\chi^2_\nu$ couples to both spin-1/2 and
spin-3/2 states.  The three vacuum sum rules obtained from the
correlator $\Pi_{\mu\nu}^{12}(q)$ at $\rho_N=0$ have been studied
extensively in Refs.~\cite{leinweber1,leinweber2}. To obtain the 
finite-density sum rules from this mixed correlator, we follow the
procedures established in Ref.~\cite{cohen1,furnstahl1,jin1,jin2,cohenr}.  
The correlator $\Pi_{\mu\nu}^{12}(q)$ contains nine distinct structures 
at finite density.  Here we only focus on the odd-chirality sum rules at 
the structures $\gamma_\mu \gamma_\nu$ and $\gamma_\mu \rlap{/}q q_\nu$, 
which are
\begin{eqnarray}
\lambda_1\, \lambda_2\, M_N^*\, e^{-(E_q^2-{\bf q}^2)/M^2} &=&
-{1 \over 8\pi^2}\, \langle\overline{q}q\rangle_{\rho_N}\, 
M^4\, E_1\, L^{8/27}
+2\, \overline{E}_q\, \langle\overline{q}q\rangle_{\rho_N}
\langle q^\dagger q\rangle_{\rho_N}\, L^{8/27}
\nonumber
\\*[7.2pt]
&&-{1+3\beta\over 96}\, \langle\overline{q}q\rangle_{\rho_N}\,
\langle{\alpha_s\over \pi}\, G^2\rangle_{\rho_N} \, L^{8/27} \, ,
\label{sum-3}
\end{eqnarray}
and
\begin{eqnarray}
\lambda_1\, \lambda_2\, {1\over M_N^*}\, e^{-(E_q^2-{\bf q}^2)/M^2}
&=& 
-{1 \over 8\pi^2} \, \langle\overline{q}q\rangle_{\rho_N}\,
M^2\, E_0\, L^{8/27}
-{3-\beta\over 64\pi^2}\,
\langle g_c \overline{q}\sigma\cdot G q\rangle_{\rho_N}\, L^{-2/9}
\nonumber
\\*[7.2pt]
&&+{1+3\beta\over 96}\, \langle\overline{q}q\rangle_{\rho_N}\,
\langle{\alpha_s\over \pi}\, G^2\rangle_{\rho_N}\, {1\over M^2}\,
L^{8/27} \, .
\label{sum-4}
\end{eqnarray}
Here $\lambda_2$ denotes the coupling of $\chi^2_{\nu}$ to the
quasi-nucleon state.  The contributions of the even-chirality
four-quark condensates to these two sum rules are multiplied 
by the current quark masses and are thus suppressed. In-medium
factorization is adopted for the odd-chirality four-quark condensates.
Here, dimension-five quark and quark-gluon condensates that vanish 
in vacuum are neglected as their contributions are numerically small \cite{jin2}.
The other sum rules obtained from the mixed correlator are either 
dependent on the even-chirality four-quark condensates or have 
vanishing Wilson coefficients for the majority of leading
order OPE terms. (For example, the Wilson coefficients for the 
operators, $\overline{q}\gamma_\mu q$, $\overline{q}\gamma_\mu D_\nu q$,
$\cdots$, $\overline{q}\gamma_\mu D_{\nu_1}\cdots D_{\nu_n} q$ vanish
in this case.)

%

In sum rules (\ref{sum-1}), ({\ref{sum-2}), (\ref{sum-3}), and
(\ref{sum-4}), we have also included the dimension-seven condensate,
$\langle\overline{q}q\rangle_{\rho_N}\, \langle{\alpha_s\over \pi}\,
G^2\rangle_{\rho_N}$, which was neglected in the previous studies.
Inclusion of its contribution should make the valid regime in Borel
mass larger and the predictions more reliable, as shown in the vacuum
sum rules \cite{leinweber2}.  As the unfactorized dimension seven
operators giving rise to $\langle\overline{q}q\rangle_{\rho_N}\,
\langle{\alpha_s\over \pi}\, G^2\rangle_{\rho_N}$ are chirally odd
their density dependence should be qualitatively similar to that of
the quark condensate.  This behavior arises naturally in the
factorized form where the density dependence of the gluon condensate
is estimated to be a 7\% effect.  There are many other dimension-seven
operators; their contributions are assumed to be relatively small
\cite{leinweber2} and are neglected.

   To analyze the sum rules, we follow the techniques introduced in
Ref.~\cite{leinweber1} for determining the valid Borel region.  We
limit the continuum model contributions to 50\% of the phenomenology,
and maintain the contributions of the highest dimensional operators in
the OPE to less than 10\% of the sum of OPE terms.  This defines a
region in Borel mass $M$ where a sum rule should be valid.  If
there is no region in which both conditions are satisfied, the sum
rule is considered to be invalid and is discarded.  

   A comparison of (\ref{sum-3}) and (\ref{sum-4}) indicates that
(\ref{sum-3}) will have a smaller region of validity than
(\ref{sum-4}).  The leading term of (\ref{sum-3}) is proportional to
$M^4$ whereas (\ref{sum-4}) is proportional to $M^2$.  The $M^4$ term
has greater overlap with excited states and limiting continuum model
contributions will restrict the upper Borel regime limit.  In
addition, a comparison of the magnitudes of the highest dimension
operators in these sum rules indicates the lower limit will be larger
for (\ref{sum-3}) when maintaining the promise of reasonable OPE
convergence.  It should not be surprising to find (\ref{sum-3}) to
become invalid prior to (\ref{sum-4}) as density increases.

   QCD sum rules relate the spectral parameters to the condensate
values and other parameters.  Any imprecise knowledge of these QCD
inputs will give rise to uncertainties in the extracted spectral
properties.  Here we follow Ref.~\cite{leinweber2} and estimate these
uncertainties via a Monte Carlo error analysis.  Gaussian
distributions for the condensate values and related parameters are
generated via Monte Carlo.  These distributions provide a distribution
for the OPE and thus uncertainty estimates that are used in the
$\chi^2$ fit. In fitting the sum rules taken from the samples of
condensate parameters, one learns how these uncertainties are mapped
into uncertainties in the extracted spectral parameters.

   As in previous works on finite-density sum rules, we use the linear
density approximation for estimating the in-medium condensates:
\begin{equation}
\langle\widehat{O}\rangle_{\rho_N}=\langle\widehat{O}\rangle_{\rm vac}
+\langle\widehat{O}\rangle_N \, \rho_N\ .
\end{equation}
The values of vacuum condensates we use are \cite{leinweber2}
$a=-4\pi^2\, \langle\overline{q}q \rangle_0=0.52\pm 0.05$ GeV${}^3$,
$b=4\pi^2\, \langle(\alpha_s/\pi)G^2 \rangle_0=1.2\pm 0.6$ GeV${}^4$,
and $m_0^2 = -\langle g_c\overline{q}\sigma\!\cdot\!{\cal G}
q\rangle_0/ \langle\overline{q}q \rangle_0=0.72 \pm 0.08$ GeV${}^2$.
The quark mass $m_q$ is chosen to satisfy the Gell-Mann--Oakes--Renner
relation, $2\,m_q\,\langle\overline{q}q\rangle_0=-m_\pi^2\, f_\pi^2$.
We adopt $\sigma_N = 0.045 \pm 0.007$ GeV \cite{gasser1},
$\langle (\alpha_s/\pi) \, G^2 \rangle_N = -0.65 \pm 0.10$ GeV, 
$\langle q^\dagger iD_0 q \rangle_{N} = 0.18 \pm 0.04$ GeV, 
$\langle g_c q^\dagger\sigma\!\cdot\!{\cal G} q\rangle_{N} = -0.1 \pm
0.5$ GeV${}^2$,
$\langle q^\dagger iD_0 iD_0 q\rangle_{N}+{1 \over 12} \langle
g_c q^\dagger\sigma\!\cdot\!{\cal G}q\rangle_{N} = 0.031 \pm 0.010$
GeV${}^2$ \cite{cohenr}.
Since the mixed condensate $\langle g_c \overline{q}\sigma\cdot G
q\rangle_{\rho_N}$ is chirally odd, we assume the same density
dependence as for the quark condensate.  We take $\mu=0.5$ GeV and
$\Lambda_{\rm QCD}= 150 \pm 40$ MeV and $\alpha_s / \pi = 0.0117 \pm
0.014$ at 1 GeV${}^2$ \cite{leinweber2}.  To explore deviations from
in-medium factorization of the chirally-even four-quark operators
contributing to the last term of (\ref{sum-2}) we introduce a 100\%
standard error and consider $\kappa = 1.0 \pm 1.0$.

   Before proceeding with the numerical analysis it is interesting to
examine the density dependence of the new sum rules of (\ref{sum-3})
and (\ref{sum-4}).  For both of these sum rules, the predominant
density dependence is governed by the quark condensate.  This density
dependence is common to all terms of these OPEs.  As such, the effects
of increasing density will be to reduce the residue of the pole while
the pole position remains largely unchanged.  This result should be
robust, and is the key result absent in the former in-medium nucleon
analysis.  The approximate invariance of $M^* + \Sigma_v$ is manifest
in (\ref{sum-4}) and is in accord with the expectations of
phenomenology.

   Our goal here is to evaluate the degree of consistency between the 
QCD sum rules and the expectations of relativistic mean-field 
phenomenology.  The one firm conclusion from previous in-medium studies 
is that the vector self-energy is positive and a few hundred MeV.  
Hence, we begin by fixing $\Sigma_v = 0.25$
GeV at saturation density and searching for a region in which both
sides of the QCD sum rules are valid \cite{leinweber2}.  These
considerations eliminate the sum rules (\ref{sum-1}) and (\ref{sum-3})
{}from the following analysis as there is no valid Borel region as
defined above.  The implications of this result will be considered in a
later work \cite{furnstahl2}.  Thus we proceed with the sum rules
(\ref{sum-2}) and (\ref{sum-4}) only.

   The parameter $\beta$ should be selected to minimize continuum
model contributions while maintaining reasonable higher-dimension
operator contributions such that the pole may be resolved from the
continuum contributions \cite{leinweber2,leinweber3}.  The sum rule
(\ref{sum-2}) indicates that these criteria cannot be satisfied
simultaneously.  We therefore select $\beta = -0.7$ for this sum rule
to suppress the continuum contribution from the third term in the OPE.
The continuum model contribution of the sum rule (\ref{sum-4}) is
independent of $\beta$.  We therefore simply set $\beta = 0$, as it is
known that the second term of (\ref{field-1}) has little overlap with
the ground state nucleon \cite{leinweber2,leinweber3}.

   Ideally, the sum rules of (\ref{sum-2}) and (\ref{sum-4}) would be
optimized by a six parameter search of the two residues, two continuum
thresholds and the scalar and vector self-energies.  However, the
final terms of (\ref{sum-2}) independent of the continuum model are
small.  A singular value decomposition of the covariance matrix
reveals that these terms are not identified as a degree of freedom.
Hence, to proceed, we have no choice but to equate the continuum
thresholds of (\ref{sum-2}) and (\ref{sum-4}).  In practice, this
approximation is acceptable when the continuum model contributions are
restricted as described above \cite{leinweber2}. 

   Fig.\ \ref{fig-1} displays the valid Borel regimes for the two sum
rules (\ref{sum-2}) and (\ref{sum-4}).  The corresponding fit for the
central values of the QCD input parameters is illustrated in Fig.\
\ref{fig-extra}.  The distributions for $M^*$ and $\Sigma_v$ are
illustrated in Fig. \ref{fig-2}.  We find $M^* = 0.70^{+0.14}_{-0.10}$
GeV and $\Sigma_v = 0.32^{+0.07}_{-0.11}$ GeV at nuclear matter
saturation density.  Normalizing these results to the vacuum nucleon
mass predicted by (\ref{sum-4}) provides our final result of $M^* =
0.64^{+0.13}_{-0.09}$ GeV and $\Sigma_v = 0.29^{+0.06}_{-0.10}$ GeV.
These results are consistent with the expectations of relativistic
phenomenology.

\begin{figure}[p]
\begin{center}
\epsfysize=11.7truecm
\leavevmode
\setbox\rotbox=\vbox{\epsfbox{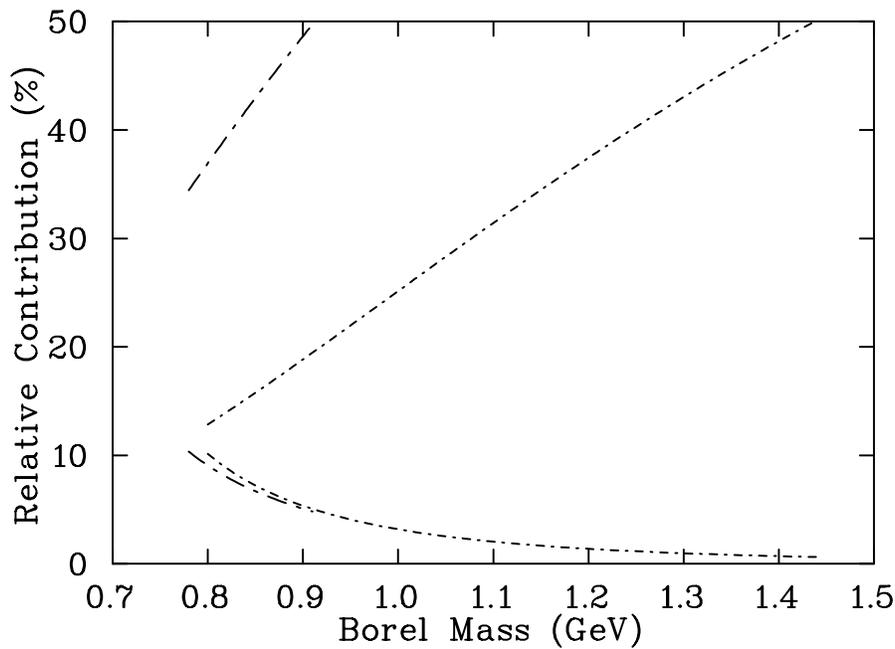}}\rotl\rotbox
\end{center}
\caption{Illustration of the valid Borel regimes for the sum rules of
(\protect\ref{sum-2}) (large dash) and (\protect\ref{sum-4}) (fine
dash).  Both continuum model contributions (limited to 50\% of the
phenomenology) and highest dimension operator contributions (limited
to 10\% of the OPE) are illustrated. }
\label{fig-1}
\end{figure}

\begin{figure}[p]
\begin{center}
\epsfysize=11.7truecm
\leavevmode
\setbox\rotbox=\vbox{\epsfbox{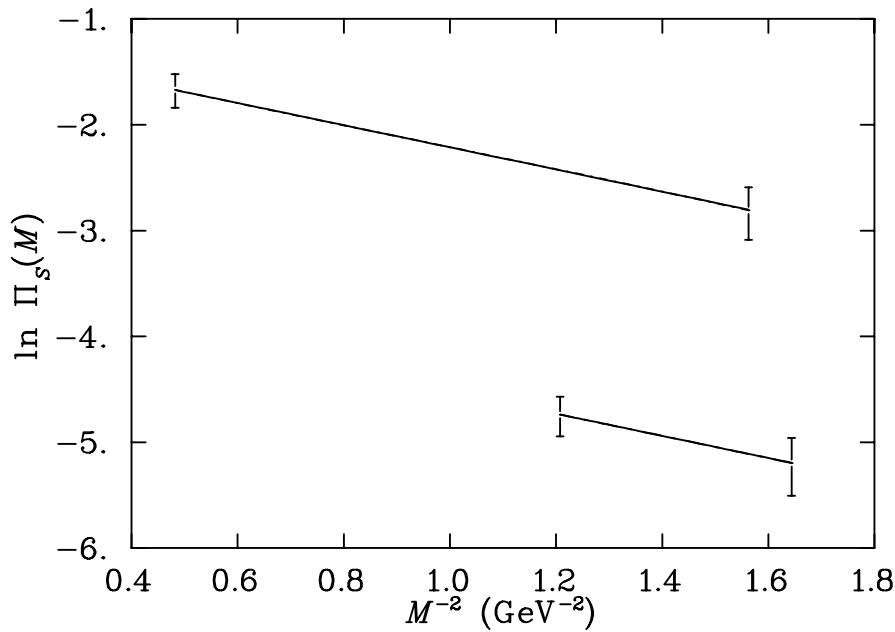}}\rotl\rotbox
\end{center}
\caption{The five parameter fit of the sum rules of
(\protect\ref{sum-2}) (large dash, lower) and (\protect\ref{sum-4})
(fine dash, upper).  The QCD$-$continuum (dashed) curves are hidden by
the ground state (solid) curves in the near perfect fits.  Only 2 of
the 51 error bars used in the $\chi^2$ fit are illustrated.}
\label{fig-extra}
\end{figure}

\begin{figure}[t]
\begin{center}
\epsfysize=11.7truecm
\leavevmode
\setbox\rotbox=\vbox{\epsfbox{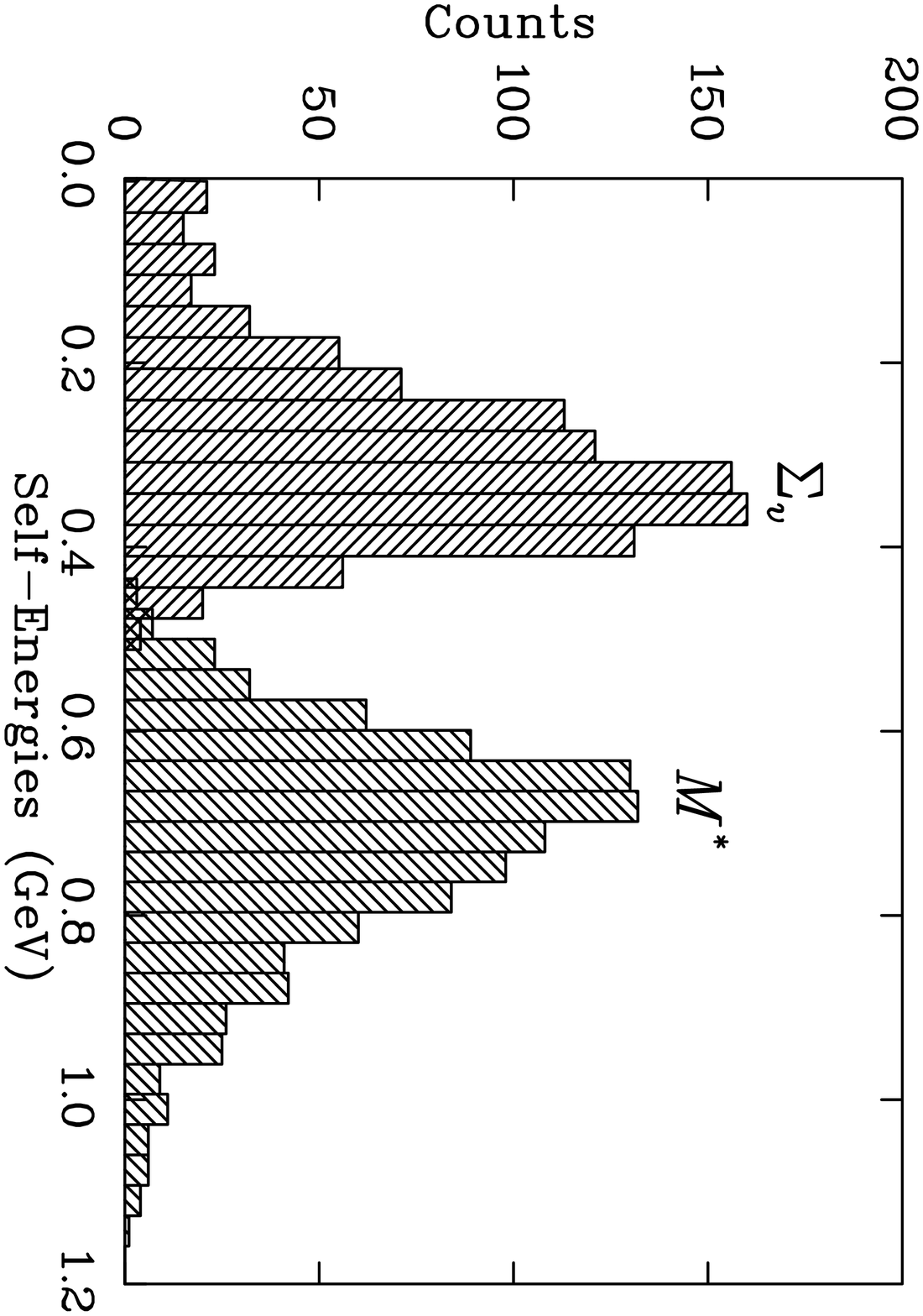}}\rotl\rotbox
\end{center}
\caption{Histogram for $\Sigma_v$ and $M^*$ obtained from a Monte
Carlo sample of 1000 QCD parameters.}
\label{fig-2}
\end{figure}

   The density dependence of these results near saturation density
appears to be weaker than previous analyses.  Table \ref{SEratios}
summarizes ratios of the finite density spectral parameters obtained
from (\ref{sum-2}) and (\ref{sum-4}) at various densities.  Median and
standard errors from the median are reported as the distributions are
not Gaussian.  The trend of the ratios is in qualitative but not
quantitative accord with the expectations of mean-field phenomenology.
The uncertainties in the ratios for $M^* + \Sigma_v$ confirm that a
determination of binding energies the order of 16 MeV is beyond
present QCD sum rule analyses.  

\begin{table}[b]
\caption{Ratios of the finite density spectral parameters at various
  densities.  Vacuum ratios report saturation-density / zero-density
  results as in $\Sigma_v / M_N$, $M^*/M_N$, $(M^* + \Sigma_v) / M_N$,
  $w^*/w$ and $\lambda_1^*\lambda_2^*/\lambda_1\lambda_2$ while
  saturation density ratios report finite-density / saturation-density
  results.}
\label{SEratios}
\begin{tabular}{cccc}
Self-Energy       &Vacuum Ratio &\multicolumn{2}{c}{Saturation Ratios} \\ 
                  &$\rho_N = 1.0$  &$\rho_N = 0.5$  &$\rho_N = 1.5$  \\
\tableline
$\Sigma_v$        &$0.30 \pm {}^{0.06}_{0.14}$ &$0.88 \pm 0.09$ &$1.09 \pm 0.07$ \\
$M^*$             &$0.69 \pm {}^{0.14}_{0.07}$ &$1.06 \pm 0.05$ &$0.96 \pm 0.05$ \\
$M^* + \Sigma_v$  &$0.99 \pm 0.03$ &$1.00 \pm 0.03$ &$1.00 \pm 0.04$ \\
$w$               &$1.01 \pm 0.04$ &$1.00 \pm 0.03$ &$1.00 \pm 0.05$ \\
$\lambda_1^2$     &                &$0.55 \pm 0.10$ &$1.43 \pm 0.29$ \\
$\lambda_1\lambda_2$   
                  &$0.46 \pm 0.08$ &$1.35 \pm 0.15$ &$0.72 \pm 0.12$ \\
%
\end{tabular}
\end{table}

   The analysis here has focused on two of the four sum rules which
are insensitive to the even-chirality four-quark condensates and hold
promise of providing reliable information on the density dependence of
spectral parameters.  Further examination of the sum rules is
necessary to determine whether the sum rules which have been found to
be invalid are at least consistent with the more reliable sum rules.

   In summary, we have obtained two new QCD sum rules for nucleons in
nuclear matter from a mixed correlator of spin-1/2 and spin-3/2
interpolating fields.  These sum rules are insensitive to the large
and poorly known even-chirality four-quark condensates and provide
additional information on the nucleon scalar self-energy in nuclear
matter.  We analyzed these new sum rules, along with the sum rules
obtained from spin-1/2 correlator.  Uncertainties of the predictions
due to imprecise knowledge of the condensate values and related
parameters were obtained from a Monte Carlo error analysis.  We found
that the QCD sum-rule predictions for nucleon self-energies are
consistent with the magnitudes but not the density dependence obtained
in relativistic nuclear phenomenology.

\acknowledgements

   R.J.F.\ acknowledges support from the US National Science
Foundation under grants PHY--9511923 and PHY--9258270 and the Sloan
Foundation.  X.J. acknowledges support from the Natural Sciences and
Engineering Research Council of Canada.  D.B.L. acknowledges support
{}from the U.S. Department of Energy under grant DE-FG06-88ER40427.


\end{document}